\documentclass[aps,twocolumn,pra,twoside,amssymb,amsmath]{revtex4}
\usepackage{amssymb}
\usepackage{graphicx}
\usepackage{amsmath}
\usepackage{colordvi}
\usepackage{bbm}
\usepackage{verbatim}
\usepackage{float}
\usepackage{dcolumn}
\usepackage[colorlinks,linkcolor=blue,anchorcolor=blue,citecolor=blue,urlcolor=blue]{hyperref}
\usepackage{mathrsfs}

\usepackage{multirow}
\usepackage{makecell}
\begin{document}
\title{Component-selective engineering of two-photon sources in a cavity-coupled two-atom system}
\author{Zhicai Chen$^{1}$, Jun Xu$^{1}$, Deyi Kong$^{2}$,  Xiangming Hu$^{1,}$\footnote{xmhu@ccnu.edu.cn} and Fei Wang$^{2,}$\footnote{feiwang@hbut.edu.cn}}
\affiliation{
  $^{1}$\mbox{College of Physical Science and Technology, Central China Normal University, Wuhan 430079, China} \\
  $^{2}$\mbox{School of Science, Hubei University of Technology, Wuhan 430068, China}}
%=============================================================================================================
\begin{abstract}
%We propose a source-oriented component-selection strategy for generating two-photon quantum light in a cavity-coupled two-atom system. For the cavity-field source, this strategy suppresses both the one- and three-photon backgrounds at comparable two-photon population, although a residual single-photon contribution remains. This residual background reflects a structural tradeoff in weakly driven, excitation-number-conserving light--matter systems: a more purely photonic target state weakens the spectral anharmonicity, whereas photon-number selectivity requires appreciable light--matter hybridization. At another operating point, selecting the double-atomic-excitation component of the same two-excitation manifold generates strongly correlated fluorescence photon pairs, enabling two distinct two-photon sources within the same physical platform. This source-oriented strategy provides a route for designing multiphoton quantum-light sources in hybrid light--matter systems.
In multiphoton blockade, drive-induced excitations are not necessarily fully converted into cavity photons, leaving unwanted lower-photon-number contributions that degrade source purity. This motivates a source-oriented component-selection strategy, in which the internal composition of the resonant two-excitation state is tailored to favor the component associated with the desired quantum-light output. In a cavity-coupled two-atom system, by engineering the resonant-state composition and controlling the interference between excitation pathways, we suppress both single- and three-photon backgrounds at matched two-photon populations, although a residual single-photon contribution remains. This residual background originates from a structural tradeoff between the photonic character of the target state and the spectral anharmonicity required for photon-number selectivity. Furthermore, selecting the double-atomic-excitation component at another operating point enables the generation of strongly correlated fluorescence photon pairs, demonstrating two distinct two-photon sources within the same physical platform.
\end{abstract}

\maketitle
\section{Introduction}
The generation of nonclassical light is central to quantum optics and quantum information science, where antibunched photons, photon-number states, and strongly correlated few-photon fields provide basic resources for quantum communication, quantum computation, and quantum metrology \cite{Kimble1977,Leonski1994,Carmichael1991,Kimble2008,Gisin2007,Ladd2010,Giovannetti2011}.
Among various mechanisms, photon blockade provides a direct route to nonclassical photon statistics by exploiting optical nonlinearity at the few-photon level \cite{Tian1992,Imamoglu1997,Werner1999,Rebic1999,Rebic2002}.
In a strongly nonlinear quantum system, the energy spectrum becomes anharmonic, so that the resonance condition for the first excitation differs from that for subsequent excitations.
As a result, when the system is resonantly driven to the single-excitation manifold, transitions to higher excitation manifolds can be off-resonant, leading to photon blockade and antibunched emission.
Conventional single-photon blockade has been extensively studied and demonstrated in various light--matter systems, including atom-cavity QED \cite{Birnbaum2005,Schuster2008}, semiconductor, polaritonic, and solid-state platforms \cite{Faraon2008,Reinhard2012,Kyriienko2020}, superconducting circuit-QED systems \cite{Lang2011,Hoffman2011,Bozyigit2011,Chakram2022}, and multi-emitter cavity-QED systems \cite{Trivedi2019,Chen2022}.
In addition to blockade based on strong intrinsic anharmonicity, interference-based unconventional photon blockade can produce strong antibunching even in weakly nonlinear coupled systems, providing another important route to single-photon generation \cite{Liew2010,Bamba2011,Flayac2017,Snijders2018,Vaneph2018,Zubizarreta2020,Hou2019}.

Beyond single-photon blockade, higher-order photon blockade aims at selectively allowing a fixed number of photons while suppressing the absorption of additional photons.
In particular, two-photon blockade corresponds to an enhanced two-photon component accompanied by the suppression of three-photon excitation, which can be characterized by $g_{\rm cav}^{(2)}(0)>1,\, g_{\rm cav}^{(3)}(0)<1$ \cite{Miranowicz2013,Hovsepyan2014,Deng2015}. Such a photon-number-selective effect has been observed in an atom-driven cavity-QED system, where the cavity output exhibits two-photon bunching together with three-photon antibunching \cite{Hamsen2017}.
On the theoretical side, higher-order photon blockade and related few-photon correlation effects have been explored in a variety of nonlinear quantum optical systems.
These include multiphoton Jaynes-Cummings and quantum Rabi systems \cite{Felicetti2018,VillasBoas2019,Zou2020,Li2024TJC,Kubanek2008}, cavity-QED systems with collective or cascaded atomic couplings \cite{Zhu2017,Bin2018,Lin2019} and Kerr-type nonlinear systems \cite{Tang2024,Zhang2025}.
In addition, interference, squeezing, nonlinear mixing, and optomechanical nonlinearities have been employed to engineer photon blockade and strongly correlated multiphoton emission \cite{Qiao2024,Kowalewska2019,Feng2021,Li2024FW,Lin2024,Rabl2011,Nunnenkamp2011,Xu2013,Liao2013,Komar2013,Xie2016,Solki2023,Tian2025Disorder}.
Nonreciprocity, non-Hermiticity, exceptional points, and dynamical control provide further routes to tailor photon statistics and blockade effects \cite{Huang2018,Li2019,Gou2023Simultaneous,Lu2024Chiral,Xu2018NRQuadratic,Xu2019Synthetic,Li2024Loss,Sun2023,Huang2022EP,Geng2024,Ghosh2019,Li2022,Zhang2024}.
%Related ideas have also been extended to correlated multiphoton-bundle emission and disorder-engineered waveguide-QED systems \cite{Wang2023TwoModeBundles,Tian2025Disorder}.
Compared with conventional single-photon blockade, higher-order photon blockade requires selective access to a target multiphoton excitation manifold while suppressing further excitation to higher manifolds. For quantum-light generation, however, selective excitation of the target manifold alone is generally not sufficient. This is already evident in weakly driven, excitation-number-conserving light--matter systems, where the dressed states can be organized into well-defined excitation manifolds. In the simplest Jaynes--Cummings case, the $N$-excitation manifold is composed of the bare states $|N,g\rangle$ and $|N-1,e\rangle$, which contain different cavity photon numbers.
For single-photon blockade, the corresponding first-excitation manifold contains the  single-photon state $|1,g\rangle$ and the zero-cavity-photon component $|0,e\rangle$.
The latter mainly reduces the weight of the desired single-photon component and hence the source brightness, but does not introduce an unwanted photon background.
For higher excitation manifolds, in contrast, the lower-photon-number components themselves constitute an unwanted photon background, thereby degrading the purity and brightness of the source.
Therefore, for higher-order photon sources, it is important not only to select the desired excitation manifold and suppress higher-order excitations, but also to selectively engineer the internal photon--matter composition of the resonant dressed state.

This consideration motivates a source-oriented component-selection strategy in a cavity-coupled two-atom system,
where a single-mode cavity interacts with two two-level atoms driven by classical fields of the
same frequency with tunable relative amplitude and phase. The detunings tailor the
composition of the resonant two-excitation state toward the desired source component, whereas
the relative driving phase controls the interference between excitation pathways. Related
two-atom cavity-QED systems have been studied for collective multiphoton blockade \cite{Zhu2017}, but our focus is on the component-selective engineering of the resonant manifold for quantum-light generation. Compared with the conventional single-atom model, the proposed strategy effectively suppresses the unwanted single-photon background, although a considerable residual background remains. This residual background reveals a structural tradeoff in the weakly driven, excitation-number-conserving systems: making the target state more purely photonic tends to weaken the spectral anharmonicity, whereas photon-number-selective excitation requires appreciable light--matter hybridization. At another operating point, selecting the double-atomic-excitation component of the same manifold generates strongly correlated fluorescence photon pairs,
thereby enabling two distinct forms of two-photon quantum-light generation within the same physical platform.

This paper is organized as follows.
In Sec.~\ref{sec2}, we introduce the cavity-coupled two-atom model and
briefly analyze the dressed-state structure in the symmetric limit.
In Sec.~\ref{sec3}, we develop the source-oriented component-selection strategy
for the cavity-field two-photon source, evaluate the resulting source performance, and identify
the resulting structural tradeoff. In Sec.~\ref{sec4}, we apply the same framework to the
double-atomic-excitation component and demonstrate strongly correlated fluorescence
photon-pair emission. Finally, Sec.~\ref{sec5} summarizes our conclusions.

\begin{figure}[t]
  \centering
  \includegraphics[scale=1.6]{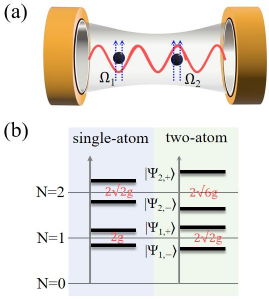}
\caption{
(a) Schematic of the cavity-coupled two-atom system.
A single cavity mode interacts with two two-level atoms, which are coherently driven by external classical fields of the same frequency with Rabi frequencies $\Omega_1$ and $\Omega_2$, respectively.
(b) Dressed-state energy ladders of the single-atom Jaynes--Cummings model (left) and the symmetric two-atom bright subspace (right).
Here, $|\Psi_{N,\pm}\rangle$ denote the upper and lower bright dressed states in the $N$-excitation manifold.
%The comparison shows the collectively enhanced dressed-state splitting and anharmonicity in the two-atom system.
}
\label{Fig1}
\end{figure}

%=============================================================================================================
\section{Model and Equations} \label{sec2}
The system under consideration is schematically shown in Fig.~\ref{Fig1}(a).
It consists of a single-mode cavity coupled to two two-level atoms.
Each atom is driven by an external classical field, and the two driving fields have the same frequency $\omega_d$ with tunable relative amplitude and phase.
In the frame rotating at the driving frequency $\omega_d$, the Hamiltonian of the system can be written as $(\hbar=1)$
\begin{equation}
H=H_0+H_{\rm int}+H_{\rm d},
\label{Eq:H}
\end{equation}
where
\begin{align} \nonumber
H_0 &= \Delta_a a^\dagger a
+\sum_{j=1}^{2}\Delta_j\sigma_j^\dagger\sigma_j, \\ \nonumber
H_{\rm int} &= \sum_{j=1}^{2}g_j\left(a\sigma_j^\dagger+a^\dagger\sigma_j\right), \\ 
H_{\rm d} &=\sum_{j=1}^{2}
\left(\Omega_j\sigma_j^\dagger+\Omega_j^*\sigma_j\right).
\end{align}
Here $a$ $(a^\dagger)$ is the annihilation (creation) operator of the cavity mode, and $\sigma_j=|g\rangle_j\langle e|$ and $\sigma_j^\dagger=|e\rangle_j\langle g|$ are the lowering and raising operators of atom $j$, respectively.
The detunings are defined as $\Delta_a=\omega_a-\omega_d$ and $\Delta_j=\omega_j-\omega_d$, where $\omega_a$ and $\omega_j$ are the cavity and atomic transition frequencies, respectively.
The parameter $g_j$ denotes the coupling strength between the cavity mode and atom $j$, and $\Omega_j$ is the complex Rabi frequency of the coherent drive applied to atom $j$.
The two complex Rabi frequencies are parametrized as
\begin{eqnarray}
\Omega_1 = \frac{\Omega}{\sqrt{1+r^2}},\,\, \Omega_2 = \frac{r\Omega e^{i\phi}}{\sqrt{1+r^2}},
\label{Eq:drive}
\end{eqnarray}
where $r$ and $\phi$ describe the relative amplitude and phase between the two driving fields.
This parametrization keeps the total driving strength fixed, namely $|\Omega_1|^2+|\Omega_2|^2=|\Omega|^2$.
In the single-atom reference model, the second atom is removed, or equivalently treated as an uncoupled spectator in its ground state. The Hamiltonian then reduces to the standard driven Jaynes-Cummings model,
\begin{eqnarray} \nonumber
H_{\rm JC} &=& \Delta_a a^\dagger a+\Delta_1\sigma_1^\dagger\sigma_1 +g_1(a\sigma_1^\dagger+a^\dagger\sigma_1) \\ &&+\Omega(\sigma_1^\dagger+\sigma_1),
\label{Eq:HJC}
\end{eqnarray}
which will be used as a reference system in the following analysis.  The dissipative dynamics of the system is described by the Lindblad master equation
\begin{eqnarray} \nonumber
\frac{d\rho}{dt} &=& -i[H,\rho]
+\frac{\kappa}{2}
\left(2a\rho a^\dagger-a^\dagger a\rho-\rho a^\dagger a\right) \\
&&+\sum_{j=1}^{2}\frac{\gamma_j}{2}
\left(2\sigma_j\rho\sigma_j^\dagger
-\sigma_j^\dagger\sigma_j\rho
-\rho\sigma_j^\dagger\sigma_j\right),
\label{Eq:master}
\end{eqnarray}
where $\kappa$ is the decay rate of the cavity mode and $\gamma_j$ is the spontaneous emission rate of atom $j$.

\begin{figure*}[t]
\centering
\includegraphics[width=\textwidth]{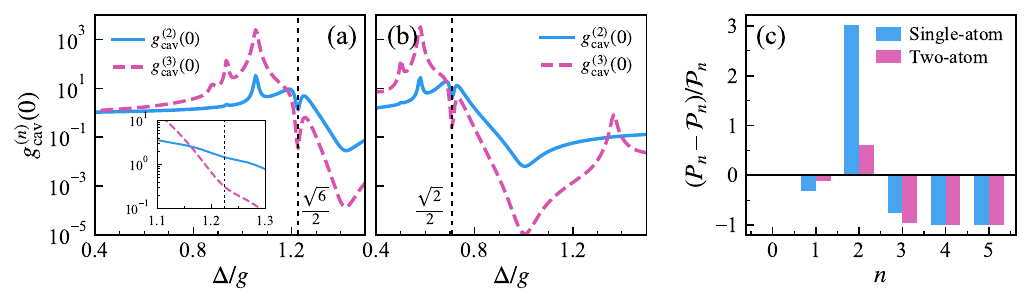}
\caption{ Equal-time correlation functions $g_{\rm cav}^{(2)}(0)$ and $g_{\rm cav}^{(3)}(0)$ as functions of the normalized detuning $\Delta/g$ for (a) two-atom system and (b) single-atom system. The inset in (a) shows the corresponding two-atom result for larger dissipation $\kappa=\gamma_1=\gamma_2=0.1g$ and stronger drive $\Omega=0.1g$.
(c) Relative deviation of the photon-number distribution from the Poisson distribution with the same mean photon number, $(P_n-\mathcal{P}_n)/\mathcal{P}_n$, at the analytical two-photon resonance points. Other parameters are $\kappa=\gamma_1=\gamma_2=0.01g$, and $\Omega=0.04g$.
For the two-atom system, the total driving strength is fixed by setting $\Omega_1=\Omega_2=\Omega/\sqrt{2}$.
}
\label{Fig2}
\end{figure*}

Before analyzing the more general detuned and phase-controlled cases, we first consider a simple symmetric limit to illustrate the dressed energy structure of the system.
We set $g_1=g_2=g$ and $\Delta_a=\Delta_1=\Delta_2=\Delta$, and temporarily neglect the coherent driving.
The Hamiltonian then reduces to
\begin{eqnarray}
H_{\rm s} &=& \Delta \hat{N}
+g\left(a\hat{S}^{+}+a^\dagger \hat{S}^{-}\right),
\label{Eq:Hs}
\end{eqnarray}
where $\hat{S}^{+} = \sigma_1^\dagger+\sigma_2^\dagger$, $\hat{S}^{-} = \sigma_1+\sigma_2$, and $\hat{N} = a^\dagger a+\sigma_1^\dagger\sigma_1+\sigma_2^\dagger\sigma_2$ is the total excitation-number operator.
Since $[H_{\rm s},\hat{N}]=0$, the Hamiltonian can be diagonalized independently in each subspace with a fixed total excitation number.

For the subspace with a fixed excitation number $N\geq 2$, the two singly excited atomic
states can be rewritten as the symmetric and antisymmetric collective states,
$|S\rangle=\frac{1}{\sqrt{2}}\bigl(|eg\rangle+|ge\rangle\bigr)$ and
$|A\rangle=\frac{1}{\sqrt{2}}\bigl(|eg\rangle-|ge\rangle\bigr)$.
The symmetric state $|S\rangle$ is the collective bright state coupled to the cavity field,
whereas the antisymmetric state $|A\rangle$ is a two-atom dark state.
Indeed, for the symmetric coupling considered here, one has
$\hat{S}^{+}|A\rangle=0$ and $\hat{S}^{-}|A\rangle=0$.
Hence the state $|N-1,A\rangle$ is decoupled from the atom--cavity interaction and has
the energy $N\Delta$.
The remaining cavity-coupled bright subspace is spanned by
${\cal B}_N=\{|N,gg\rangle,|N-1,S\rangle,|N-2,ee\rangle\}$.
Thus, in the basis ${\cal B}_N$, the Hamiltonian is written as
\begin{equation}
H_N=N\Delta I+g
\begin{pmatrix}
0 & \sqrt{2N} & 0 \\
\sqrt{2N} & 0 & \sqrt{2(N-1)} \\
0 & \sqrt{2(N-1)} & 0
\end{pmatrix},
\label{Eq:HN}
\end{equation}
where $I$ is the $3\times3$ identity matrix.
The three dressed energies in the bright subspace of the $N$-excitation manifold are
$E_{N,0}=N\Delta$ and $E_{N,\pm}=N\Delta\pm g\sqrt{4N-2}$.
For comparison, in the corresponding single-atom Jaynes--Cummings model,
the $N$-excitation manifold is spanned by
$\{|N,g\rangle,|N-1,e\rangle\}$, and the dressed energies are
$E_{N,\pm}^{(1)}=N\Delta\pm g\sqrt{N}$.
The dressed-state splittings are
\begin{equation}
\Delta E_N^{(2)}=2g\sqrt{4N-2},\qquad
\Delta E_N^{(1)}=2g\sqrt{N}.
\label{Eq:splitting}
\end{equation}
For $N=1$, the bright subspace is two-dimensional and the splitting is
$\Delta E_1^{(2)}=2g\sqrt{2}$.
For all relevant excitation manifolds,
$\Delta E_N^{(2)}>\Delta E_N^{(1)}$, showing that the collective coupling of two atoms
produces a larger dressed-state splitting than the single-atom case.
Figure~\ref{Fig1}(b) schematically compares the single-atom Jaynes--Cummings ladder
with the symmetric two-atom bright-state ladder, where $|\Psi_{N,\pm}\rangle$
denote the upper and lower dressed states with eigenenergies $E_{N,\pm}$.
This collective enhancement provides a useful spectral basis for engineering
two-photon excitation and photon blockade in the two-atom system.

%==========================================================================================================

\section{Enhanced cavity-field two-photon blockade} 
\label{sec3}
Based on the dressed energy structure established above, we now investigate the two-photon blockade effect of the cavity field. The collective two-atom coupling increases the dressed-state splitting and provides a useful spectral basis for photon-number-selective excitation.
To quantitatively characterize the photon statistics of the cavity field, we calculate the equal-time $n$th-order correlation function
\begin{eqnarray}
g_{\rm cav}^{(n)}(0) &=&
\frac{\langle a^{\dagger n}a^n\rangle}
{\langle a^\dagger a\rangle^n},
\quad n=2,3 .
\label{Eq:gcav}
\end{eqnarray}
For the two-photon blockade considered here, the relevant criterion is
\begin{eqnarray}
g_{\rm cav}^{(2)}(0)>1,\qquad g_{\rm cav}^{(3)}(0)<1 .
\label{Eq:TPBcriterion}
\end{eqnarray}
The first condition indicates the enhanced two-photon correlation, while the second one indicates the suppression of three-photon excitation.
To clarify the excitation picture, the transition from the ground state to the
two-excitation manifold is mediated by the one-excitation manifold,
$|G\rangle\rightarrow|\Psi_{1,\mu}\rangle\rightarrow|\Psi_{2,\alpha}\rangle$.
The one-excitation manifold is not resonant with a single driving photon and acts as an off-resonant intermediate manifold, whereas the total energy of two driving photons matches the  two-excitation manifold \cite{Zhu1987,Lewenstein1990,Kubanek2008}.
In the rotating frame, this two-photon resonance corresponds to
$E_{2,\alpha}=0$. We now apply this resonance condition to the dressed-state spectrum.
For the single-atom Jaynes--Cummings model, the lower dressed state in the two-excitation manifold becomes resonant when
\begin{eqnarray}
E_{2,-}^{(1)} &=& 2\Delta-\sqrt{2}g=0,
\label{Eq:res_single}
\end{eqnarray}
which gives $\Delta_{\rm 2PB}^{(1)} = \frac{\sqrt{2}}{2}g$.
Similarly, for the symmetric two-atom system, the two-photon resonance condition is
\begin{eqnarray}
E_{2,-}^{(2)} &=& 2\Delta-\sqrt{6}g=0,
\label{Eq:res_two}
\end{eqnarray}
leading to the resonant detuning  $\Delta_{\rm 2PB}^{(2)} = \frac{\sqrt{6}}{2}g$.

To verify the above analytical resonance conditions, we numerically solve the master equation and calculate the steady-state correlation functions of the cavity field.
As shown in Fig.~\ref{Fig2}(a), for the symmetric two-atom system, the correlation functions exhibit a pronounced resonance feature around the analytical position $\Delta_{\rm 2PB}^{(2)}=\sqrt{6}g/2$.
In this region, the simultaneous occurrence of $g_{\rm cav}^{(2)}(0)>1$ and $g_{\rm cav}^{(3)}(0)<1$ indicates the emergence of cavity-field two-photon blockade.
For comparison, Fig.~\ref{Fig2}(b) shows the corresponding result for the single-atom Jaynes--Cummings model, where the resonance feature appears around $\Delta_{\rm 2PB}^{(1)}=\sqrt{2}g/2$, in agreement with the dressed-state prediction.

We further comment on the role of dissipation in Fig.~\ref{Fig2}(a).
The main panels are calculated for weak dissipation, $\kappa=\gamma_1=\gamma_2=0.01g$, in order to resolve the narrow dressed-state resonances associated with the analytical two-photon resonance conditions.
The inset of Fig.~\ref{Fig2}(a) shows the corresponding two-atom result for a larger decay rate, $\kappa=\gamma_1=\gamma_2=0.1g$.
In this case, the linewidths of the dressed resonances are broadened, and the sharp spectral structures in $g_{\rm cav}^{(2)}(0)$ and $g_{\rm cav}^{(3)}(0)$ are smeared out.
Nevertheless, for the parameters shown in the inset, the correlation functions still satisfy the two-photon-blockade criterion, $g_{\rm cav}^{(2)}(0)>1$ and $g_{\rm cav}^{(3)}(0)<1$, in the vicinity of the two-photon resonance.
Thus, the weak dissipation used in the main panels should be understood as a convenient choice for identifying the dressed-state resonance positions and for comparing the single-atom and two-atom systems under identical conditions.
It is not a prerequisite for the occurrence of two-photon blockade, although the relevant linewidths must remain sufficiently narrow compared with the dressed-state anharmonicity.
Indeed, two-photon blockade has been observed experimentally in an atom-driven cavity-QED system under realistic dissipative conditions \cite{Hamsen2017}.
To further characterize the photon-number statistics, Fig.~\ref{Fig2}(c) compares the steady-state photon-number distribution $P_n$ with the Poisson distribution $\mathcal{P}_n$ having the same mean photon number by plotting the relative deviation $(P_n-\mathcal{P}_n)/\mathcal{P}_n$.
The positive deviation at $n=2$ indicates the enhancement of the two-photon component relative to the Poisson distribution, while the negative deviations at higher photon numbers show the suppression of higher-photon excitations.
This photon-number distribution, together with the correlation functions shown above, provides further evidence for the two-photon blockade effect.

\begin{figure*}[t]
\centering
\includegraphics[width=\textwidth]{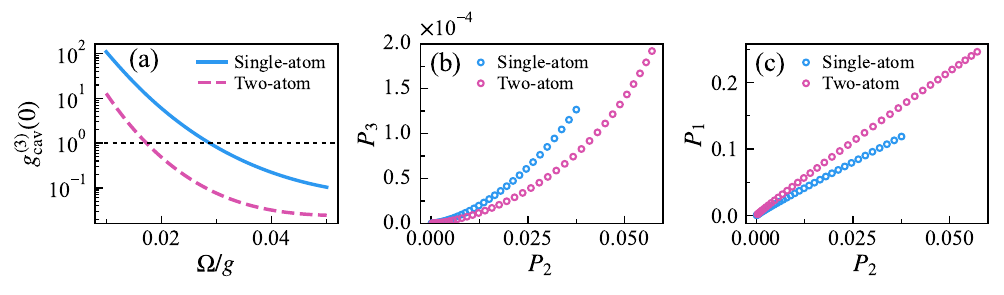}
\caption{ (a) Third-order correlation function $g_{\rm cav}^{(3)}(0)$ as a function of the normalized total driving strength $\Omega/g$.
(b) Three-photon probability $P_3$ as a function of the two-photon probability $P_2$.
(c) Single-photon probability $P_1$ as a function of $P_2$.
The data in (b) and (c) are obtained from the same drive-strength scan as in (a).
For the single-atom system, $\Delta=\sqrt{2}g/2$; for the symmetric two-atom system, $\Delta=\sqrt{6}g/2$ and $\Omega_1=\Omega_2=\Omega/\sqrt{2}$.
Other parameters are $\kappa=\gamma_1=\gamma_2=0.01g$.
}
\label{Fig3}
\end{figure*}

To gain further insight, we compare the single-atom and symmetric two-atom systems by scanning the total driving strength. As shown in Fig.~\ref{Fig3}(a), the value of $g_{\rm cav}^{(3)}(0)$ decreases with increasing driving strength in both cases, while the two-atom system exhibits a much smaller $g_{\rm cav}^{(3)}(0)$ than the single-atom system. This behavior is consistent with the larger dressed-state splitting discussed above and shows that collective two-atom coupling is beneficial for suppressing three-photon excitation. To make a fair comparison at the same two-photon population , we plot $P_3$ and $P_1$ as functions of $P_2$ in Figs.~\ref{Fig3}(b) and \ref{Fig3}(c), respectively. Figure~\ref{Fig3}(b) shows that, for a given $P_2$, the symmetric two-atom system has a smaller $P_3$ than the single-atom system, confirming the improved suppression of three-photon leakage. However, Fig.~\ref{Fig3}(c) shows that the single-photon probability $P_1$ in the symmetric two-atom system is larger than that in the single-atom case at the same $P_2$. This larger single-photon background can be understood from the detuning of the lower one-excitation branch from the driving field. At the respective two-photon resonances, the single-atom system gives $\delta_{1,-}^{(1)}=(\sqrt{2}/2-1)g$, whereas the symmetric two-atom system gives $\delta_{1,-}^{(2)}=(\sqrt{6}/2-\sqrt{2})g$. Since $|\delta_{1,-}^{(2)}|<|\delta_{1,-}^{(1)}|$, the lower one-excitation branch is closer to resonance in the symmetric two-atom system, resulting in a larger single-photon background.
These results show that collective enhancement alone does not guarantee an improved two-photon source: stronger suppression of higher-order excitation can be accompanied by increased lower-photon-number contamination. This motivates a source-oriented design beyond the symmetric configuration, in which the internal composition of the resonant two-excitation state is tailored toward the desired source component. For the cavity-field source, we therefore seek a resonant dressed state with a large $|2,gg\rangle$ component and first analyze the two-excitation manifold in the absence of the weak coherent drive.
In the basis
\begin{equation}
  \mathcal{B}_2 = \{|2,gg\rangle,\ |1,eg\rangle,\ |1,ge\rangle,\ |0,ee\rangle\},
\end{equation}
the Hamiltonian in this manifold can be written as
\begin{equation}
  H_2 =
  \begin{pmatrix}
    2\Delta_a & \sqrt{2}g_1 & \sqrt{2}g_2 & 0 \\
    \sqrt{2}g_1 & \Delta_a+\Delta_1 & 0 & g_2 \\
    \sqrt{2}g_2 & 0 & \Delta_a+\Delta_2 & g_1 \\
    0 & g_2 & g_1 & \Delta_1+\Delta_2
  \end{pmatrix}.
  \label{Eq:H2_general}
\end{equation}
A two-photon resonance occurs when an eigenenergy of $H_2$ becomes degenerate with the ground state. Since the ground-state energy is zero in our rotating frame, this condition is simply 
\begin{eqnarray}
    \det(H_2) = 0
\end{eqnarray}

\begin{figure*}[t]
\centering
\includegraphics[width=\textwidth]{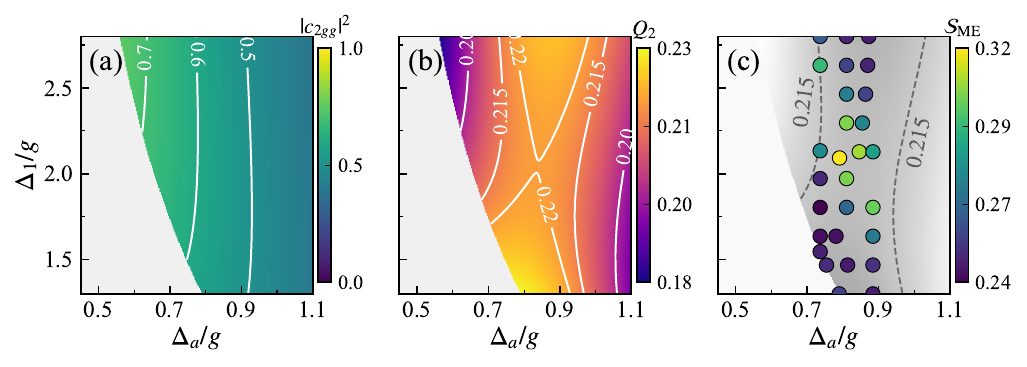}
\caption{
source-oriented component-selection strategy procedure for the cavity-field two-photon source.
For each point in the $(\Delta_a,\Delta_1)$ plane, the two-photon resonance condition $\det(H_2)=0$ is used to determine $\Delta_2$, and the resonant branch with the larger two-cavity-photon weight is selected.
Blank regions indicate parameter points for which the resonance condition gives no real solution of $\Delta_2$ within the chosen range $0 \leq \Delta_2/g \leq 4$; solutions outside this range correspond to far-detuned branches and are not retained.
(a) Two-cavity-photon weight $|c_{2gg}|^2$ of the resonant eigenstate as a function of detunings $\Delta_a$ and $\Delta_1$.
(b) Static screening indicator $Q_2$, which combines the desired two-cavity-photon weight with the intermediate-state admixture relevant to population transfer.
(c) Full master-equation validation of representative candidate points selected from the high-$Q_2$ region.
The color of each point represents $\mathcal{S}_{\rm ME}=\langle P_2/(P_1+P_3)\rangle$, evaluated over the fixed two-photon population window $0.005\le P_2\le 0.015$ under the conditions $g_{\rm cav}^{(2)}(0)>1$ and $g_{\rm cav}^{(3)}(0)<1$.
In the master-equation calculation, balanced out-of-phase driving with $r=1$ and $\phi=\pi$ is used.
}
\label{Fig4}
\end{figure*}

For simplicity, we take $g_1=g_2=g$ in the following analysis.
For given $\Delta_a$ and $\Delta_1$, the resonance condition $\det(H_2)=0$ gives a quadratic equation for $\Delta_2$,
\begin{eqnarray}
{\cal A}\Delta_2^2+{\cal B}\Delta_2+{\cal C} &=& 0,
\label{Eq:Delta2_equation}
\end{eqnarray}
where
\begin{align}
  \mathcal{A} &= \Delta_a^2 + \Delta_a\Delta_1 - g^2, \nonumber\\
  \mathcal{B} &= \Delta_a(\Delta_a+\Delta_1)^2 - g^2(3\Delta_a+2\Delta_1), \nonumber\\
  \mathcal{C} &= \Delta_a^2\Delta_1(\Delta_a+\Delta_1) - g^2(2\Delta_a^2+3\Delta_a\Delta_1+\Delta_1^2).
  \label{Eq:ABC}
\end{align}
For $\mathcal{A}\neq0$, the resonant detuning of atom 2 is
\begin{equation}
  \Delta_2 = \frac{-\mathcal{B} \pm \sqrt{\mathcal{B}^2 - 4\mathcal{A}\mathcal{C}}}{2\mathcal{A}}.
  \label{Eq:Delta2_res}
\end{equation}
The corresponding resonant eigenstate can be written as
\begin{align}
  |\psi_{\rm T}\rangle &= c_{2gg}|2,gg\rangle + c_{1eg}|1,eg\rangle \nonumber\\
  &+ c_{1ge}|1,ge\rangle + c_{0ee}|0,ee\rangle,
  \label{Eq:target_state}
\end{align}
Here, $|2,gg\rangle$ is the desired source component, and the resonant state should therefore contain a sizable weight $|c_{2gg}|^2$.
To implement this component-selective design, we use the parameter-selection procedure summarized in Fig.~\ref{Fig4}.
The purpose of this procedure is to reduce the three-dimensional detuning space $(\Delta_a,\Delta_1,\Delta_2)$ to a small set of physically motivated candidates before performing the full driven--dissipative calculation.

We first use the undriven two-excitation Hamiltonian $H_2$, obtained from $H_0+H_{\rm int}$ in the two-excitation manifold, to locate the two-photon resonant branches.
For given $\Delta_a$ and $\Delta_1$, the resonance condition $\det(H_2)=0$ determines one or more possible values of $\Delta_2$.
Among the allowed resonant branches, we select the one with the larger two-cavity-photon weight $|c_{2gg}|^2$.
In the maps, only real solutions within the practical range $0\le \Delta_2/g\le 4$ are retained; parameter points without such a solution are left blank.
Figure~\ref{Fig4}(a) shows the resulting weight $|c_{2gg}|^2$ of the selected resonant eigenstate.
A large value of $|c_{2gg}|^2$ indicates that the resonant state has a sizable projection onto $|2,gg\rangle$, providing the spectral basis for enhancing the cavity-field two-photon component.

However, a large $|c_{2gg}|^2$ alone does not ensure efficient population of this state during the full driven--dissipative dynamics.
Although several excitation pathways can contribute, the final conversion to the two-cavity-photon component must involve the intermediate states $|1,eg\rangle$ and $|1,ge\rangle$, which are directly coupled to $|2,gg\rangle$ through $|1,eg\rangle \xrightarrow{\sqrt{2}g_1} |2,gg\rangle$ and $|1,ge\rangle \xrightarrow{\sqrt{2}g_2} |2,gg\rangle$.
Therefore, a resonant state that is nearly a pure $|2,gg\rangle$ state is not necessarily favorable, because its weak admixture with $|1,eg\rangle$ and $|1,ge\rangle$ provides only limited intermediate channels for the atom-driven excitation.
On the other hand, excessive single-cavity-photon admixture would reduce the desired two-photon character of the target state.
We therefore seek a resonant state that combines a sizable $|2,gg\rangle$ component with sufficient $|1,eg\rangle$ and $|1,ge\rangle$ admixtures for population transfer.

We introduce the static screening indicator

\begin{equation}
Q_2 = |c_{2gg}|^2 W_{1c},
\label{Eq:Q2}
\end{equation}
with
\begin{equation}
W_{1c} = |c_{1eg}|^2 + |c_{1ge}|^2 .
\label{Eq:Wsingle}
\end{equation}
Here $W_{1c}$ measures the atom--cavity admixture connecting the atomic driving channels to the two-cavity-photon component. Thus, $Q_2$ provides a simple static measure balancing these two requirements. We emphasize that $Q_2$ is only a static screening indicator, not a transition rate or a final dynamical figure of merit.

The high-$Q_2$ region in Fig.~\ref{Fig4}(b) is then used to select candidate detunings for full master-equation validation.
In practice, we retain representative points satisfying $|c_{2gg}|^2>0.50$ and $Q_2>0.215$.
This static selection reduces the parameter space while leaving the effects of the
driving phase, dissipation, and leakage to higher-excitation manifolds to be tested
dynamically.
As indicated by Fig.~\ref{Fig3}, unwanted single-photon excitation can
remain sizable even when the target two-excitation manifold is resonantly selected.
We therefore use the relative phase of the two atomic drives to control the excitation
pathways. In the driven--dissipative validation, balanced out-of-phase driving with
$r=1$ and $\phi=\pi$ is employed to induce destructive interference between the two
atomic excitation pathways, thereby suppressing the dominant single-photon background
while retaining access to the target two-excitation state.
The total driving strength is scanned over $0.001\le\Omega/g\le0.2$ for each candidate detuning.
Only points satisfying the two-photon-blockade criteria
$g_{\rm cav}^{(2)}(0)>1$ and $g_{\rm cav}^{(3)}(0)<1$ are retained.
The driven--dissipative performance of the candidate points is then
evaluated from the steady-state photon-number distribution rather than
from the static indicator $Q_2$.
To compare different candidates at the same two-photon brightness, we evaluate
\begin{equation}
\mathcal{S}_{\rm ME}
=
\left\langle
\frac{P_2}{P_1+P_3}
\right\rangle ,
\label{Eq:SME}
\end{equation}
where $P_n$ is the steady-state probability of having $n$ photons in the cavity.
For a full $\Omega$ scan, we first collect all valid data points satisfying the two-photon-blockade criteria.
For each candidate whose valid data cover the target two-photon population window $0.005\le P_2\le0.015$, we interpolate $P_1$ and $P_3$ as functions of $P_2$ and evaluate $P_2/(P_1+P_3)$ on a uniform $P_2$ grid within this window.
The average over this grid gives a single figure of merit for each candidate detuning.
This matched-$P_2$ procedure allows different detuning candidates to be compared at comparable two-photon brightness.
The quantity $\mathcal{S}_{\rm ME}$ therefore measures the relative weight of the desired two-photon population against the unwanted single-photon background and three-photon leakage.

\begin{figure*}[t]
\centering
\includegraphics[width=\textwidth]{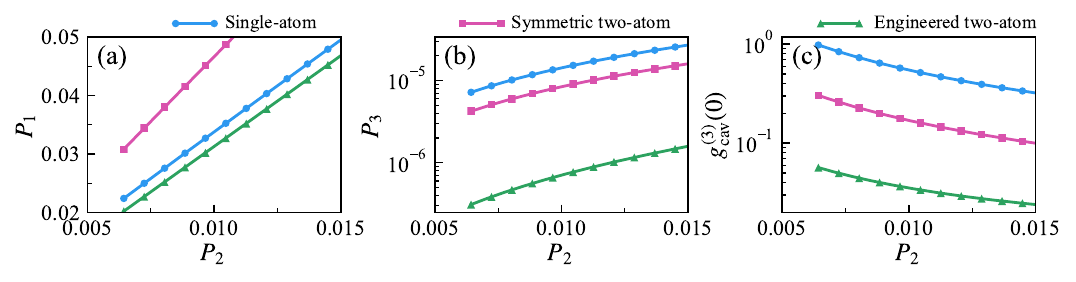}
\caption{ (a) Single-photon probability $P_1$, (b) three-photon probability $P_3$, and (c) third-order correlation function $g_{\rm cav}^{(3)}(0)$ as functions of the two-photon probability $P_2$ for the single-atom, symmetric two-atom, and engineered two-atom systems.
The curves are compared at the same $P_2$ by adjusting the driving strength in each case. The single-atom and symmetric two-atom systems are evaluated at their analytical two-photon resonances, $\Delta=\sqrt{2}g/2$ and $\Delta=\sqrt{6}g/2$, respectively, with $\Omega_1=\Omega_2=\Omega/\sqrt{2}$ for the symmetric two-atom case.
The  engineered  two-atom system uses $\Delta_a=0.79g$, $\Delta_1=2.09g$, $\Delta_2=2.30g$, $r=1$, and $\phi=\pi$. Other parameters are $\kappa=\gamma_1=\gamma_2=0.01g$.}
\label{Fig5}
\end{figure*}

Figure~\ref{Fig4}(c) shows the resulting $\mathcal{S}_{\rm ME}$ for representative candidates selected from the high-$Q_2$ region.
The gray background is the static $Q_2$ guide, while the colored points are the full master-equation results. Larger value of $\mathcal{S}_{\rm ME}$ correspond to better two-photon source performance.
Following this procedure, we choose the representative working point
$\Delta_a=0.79g$, $\Delta_1=2.09g$, and $\Delta_2=2.30g$.
As shown below, the selected configuration suppresses both the single-photon background and the three-photon leakage while maintaining the desired two-photon population.
%============================================================================================================

The full dissipative results are shown in Fig.~\ref{Fig5}.
To make a fair comparison, we use the two-photon probability $P_2$ as the horizontal axis.
For each system, the driving strength is varied separately, and the resulting single-photon background and three-photon leakage are compared at the same target two-photon population .
Figure~\ref{Fig5}(a) shows that the symmetric two-atom system has a larger single-photon probability $P_1$ than the single-atom system, consistent with the limitation identified above.
In contrast, the engineered two-atom system reduces $P_1$ below the single-atom result over the whole displayed range of $P_2$.
At the same time, Figs.~\ref{Fig5}(b) and \ref{Fig5}(c) show that the engineered two-atom system also exhibits the smallest three-photon probability $P_3$ and the lowest third-order correlation function $g_{\rm cav}^{(3)}(0)$.
These results demonstrate that the engineered two-atom configuration can simultaneously suppress the one- and three-photon components while maintaining the matched two-photon population $P_2$.
This improvement is also consistent with the component selection in the resonant two-excitation state, where the two-cavity-photon weight increases from $0.50$ for the single-atom system and $0.333$ for the symmetric two-atom system to $0.591$ for the engineered configuration.
The comparison of $P_1$ at the same $P_2$ provides a useful measure of the
relative single-photon background between different configurations. However,
$P_1$ alone does not quantify the actual isolated single-photon emission, because
a substantial part of the  single-photon population can arise naturally from the
decay cascade of the two-photon state,
$|2\rangle\rightarrow|1\rangle\rightarrow|0\rangle$. To directly characterize
the emitted photons, we therefore perform Monte Carlo quantum-trajectory
simulations and statistically analyze the cavity-photon emission events.

\begin{figure*}[t]
  \centering
  \includegraphics[scale=0.9]{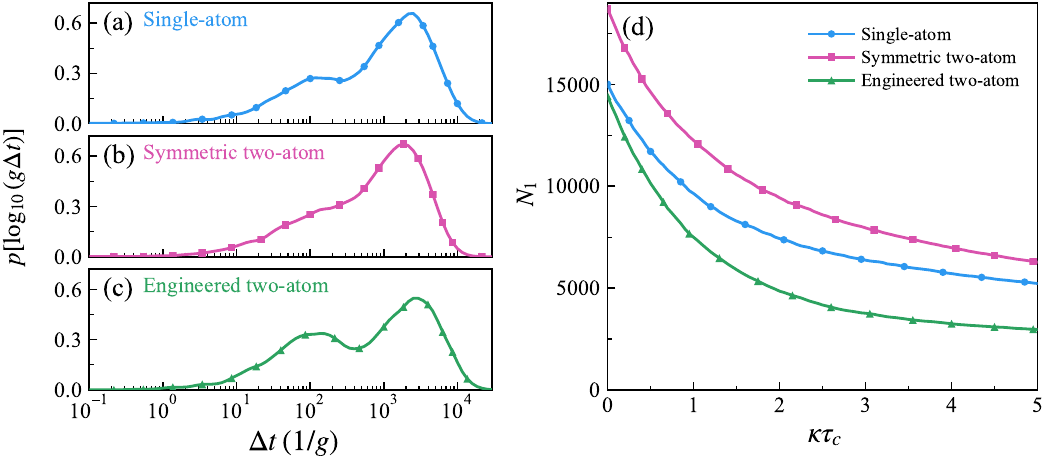}
\caption{
(a)--(c) Probability-density distributions of the interval time $\Delta t$
between consecutive cavity-photon emissions for the single-atom, symmetric
two-atom, and engineered two-atom configurations, respectively.
(d) Number of isolated single-photon emission events $N_1$ as a function
of the correlation window $\kappa\tau_c$ for the three configurations,
evaluated over the same observation time. The corresponding driving strengths are $\Omega/g=0.0326132$, $0.0248842$, and $0.1259514$ for the single-atom, symmetric two-atom, and engineered
two-atom configurations, respectively, chosen such that $P_2=0.01$.
All other parameters are the same as in Fig.~\ref{Fig5}.}
\label{Fig6}
\end{figure*}

The resulting cavity-output statistics are shown in Fig.~\ref{Fig6}.
For each trajectory, the time interval between two consecutive cavity-photon emissions is defined as $\Delta t=t_{j+1}-t_j$, where $t_j$ and $t_{j+1}$ denote their respective emission times.
To resolve the widely separated emission time scales, we introduce the logarithmic time-interval variable $x=\log_{10}(g\Delta t)$ and calculate its normalized probability density $p(x)$, with $\int p(x)dx=1$.
Figures~\ref{Fig6}(a)--\ref{Fig6}(c) show $p[\log_{10}(g\Delta t)]$ for the single-atom, symmetric two-atom, and engineered two-atom configurations, respectively.
 
The short-time peak, on the scale of $1/\kappa$, originates mainly from the time interval between the two cavity photons forming a photon pair, whereas the long-time peak reflects the much slower reexcitation process between successive photon-pair emissions.
As seen in Fig.~\ref{Fig6}(c), the larger area in the short-time region indicates that a larger fraction of consecutive cavity photons are emitted close together in time, corresponding to more pronounced photon-pair emission.

Based on this separation of time scales, we introduce a correlation window $\tau_c$ to identify correlated cavity-photon emissions. If the time interval between two consecutive cavity photons does not exceed $\tau_c$, the two photons are regarded as correlated, whereas a cavity photon without a neighboring emission within this time window is counted as an isolated single-photon event.
Figure~\ref{Fig6}(d) shows the resulting single-photon event number $N_1$ as a function of $\kappa\tau_c$.
Because there is no unique cutoff separating the two characteristic time scales, $\tau_c$ is varied to examine the robustness of the event classification. Since the three configurations are compared over the same observation time and at the same two-photon population $P_2$, the resulting $N_1$ provides a direct comparison of their isolated single-photon emission.
The engineered two-atom configuration exhibits substantially fewer isolated single-photon events over the entire displayed range of $\tau_c$. For example, at $\kappa\tau_c=3$, $N_1$ decreases from $6373$ and $7957$ for the single-atom and symmetric two-atom configurations, respectively, to $3765$ for the engineered configuration, corresponding to reductions of approximately $41\%$ and $53\%$. These results show that selecting the two-excitation manifold alone does not eliminate isolated single-photon emission, whereas the proposed component-selection strategy provides further suppression of such events.

Although the engineered two-atom system substantially reduces isolated single-photon emission events, a residual single-photon contribution remains, reflecting a structural constraint of weakly driven, excitation-number-conserving light–matter systems. This can be illustrated most simply by the single-emitter
Jaynes--Cummings model. For a high-purity two-photon source, the $|2,g\rangle$ component of the resonant
two-excitation state should strongly dominate over the lower-photon-number component $|1,e\rangle$.
However, if the two-excitation state is dominated by the bare $|2,g\rangle$ component, it becomes
increasingly cavity-like, and the corresponding energy ladder approaches that of a harmonic cavity,
reducing the spectral anharmonicity required to distinguish the two-photon resonance from neighboring
excitation manifolds. Conversely, a sufficiently anharmonic spectrum requires appreciable light--matter
hybridization, which inevitably introduces a finite and often sizable $|1,e\rangle$ component into the
target state. Thus, weakly driven, excitation-number-conserving light--matter systems exhibit a structural
tradeoff between the photonic Fock-state character of the target state and the spectral anharmonicity
required for photon-number-selective excitation. In strongly driven dressed-state schemes, by contrast, the atom is first dressed by the strong driving field, and the cavity can then be selectively excited to a target $N$-photon dressed state with a definite cavity photon number $N$, rather than a mixture of different photon-number components; for example, a two-photon bundle purity close to $100\%$ has been reported \cite{SanchezMunoz2014}.

\begin{figure}[t]
  \centering
  \includegraphics[scale=0.85]{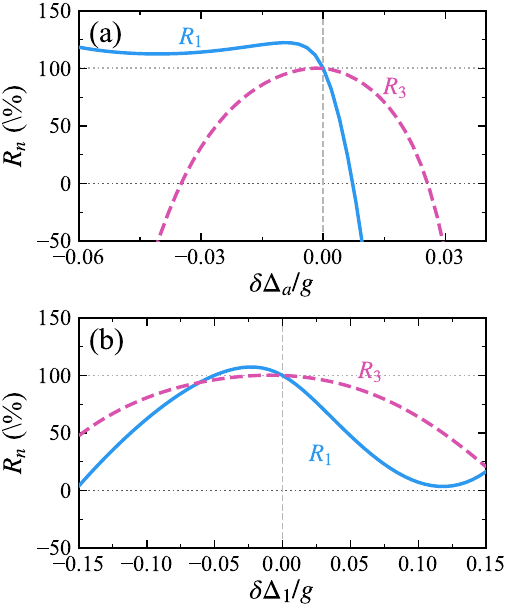}
\caption{
Robustness against detuning variations. The detuning deviations
$\delta\Delta_a$ and $\delta\Delta_1$ from the selected working point are
varied in panels (a) and (b), respectively. At each detuning deviation,
the driving strength is adjusted to maintain the same two-photon
population $P_2=0.01$, and the suppression of the single- and three-photon
backgrounds is characterized by $R_n=(P_n^{\rm SA}-P_n)/\Delta P_n$ ($n=1,3$). Here, $\Delta P_1=2.5123\times10^{-3}$ and $\Delta P_3=1.3531\times10^{-5}$ are the corresponding suppression gains at the selected working point in Fig.~\ref{Fig5}.  Other parameters are the same as in Fig.~\ref{Fig5}.}
\label{Fig7}
\end{figure}

To quantify the robustness against detuning variations, we define the normalized suppression factor as
\begin{equation}
R_n=\frac{P_n^{\rm SA}-P_n}{\Delta P_n},\qquad n=1,3,
\end{equation}
where $P_n^{\rm SA}$ denotes the photon-number population of the single-atom reference
at $P_2=0.01$, with $P_1^{\rm SA}=3.3804\times10^{-2}$ and
$P_3^{\rm SA}=1.4239\times10^{-5}$. The numerator $P_n^{\rm SA}-P_n$
measures the remaining suppression relative to the single-atom reference, while
$\Delta P_1=2.5123\times10^{-3}$ and
$\Delta P_3=1.3531\times10^{-5}$ are the corresponding suppression improvements at the
selected working point in Fig.~\ref{Fig5}. With this normalization, $R_n=1$ at the
selected working point, whereas $R_n>0$ indicates that the suppression advantage over
the single-atom reference is retained. For each detuning variation, the driving strength
is readjusted to maintain $P_2=0.01$, so that the comparison is performed at the same
two-photon population .

As shown in Fig.~\ref{Fig7}(a), both the one- and three-photon suppression advantages
are retained for approximately
$-0.04g\lesssim\delta\Delta_a\lesssim0.01g$.
In Fig.~\ref{Fig7}(b), both advantages persist throughout the investigated range
$-0.15g\le\delta\Delta_1\le0.15g$.
Thus, the engineered two-atom scheme is relatively sensitive to variations of
$\Delta_a$, while showing considerably greater robustness against variations of
$\Delta_1$.

In summary, the source-oriented component-selection strategy provides a  route to improving the performance of the cavity-field two-photon source. Within this framework, the detunings control the composition of the resonant two-excitation eigenstate, whereas the relative driving phase controls the excitation pathways through interference. The full driven--dissipative results confirm that this component-selection strategy leads to a clear improvement in the two-photon source. Nevertheless, a considerable residual background remains because of the structural tradeoff identified above between the photonic Fock-state character of the target state and the spectral anharmonicity required for photon-number-selective excitation.

\section{Strongly Correlated Fluorescence Photon Pairs}
\label{sec4}
Having characterized the cavity-field two-photon source, we now examine the atomic fluorescence and show that strongly correlated photon pairs can also be generated in the same cavity-coupled two-atom system.
To characterize the fluorescence statistics, we introduce the normalized second-order correlation function
\begin{equation}
g_{ij}^{(2)}(\tau)
=
\frac{
\langle \sigma_i^\dagger(0)\sigma_j^\dagger(\tau)\sigma_j(\tau)\sigma_i(0) \rangle
}{
\langle \sigma_i^\dagger\sigma_i\rangle
\langle \sigma_j^\dagger\sigma_j\rangle
},
\,\,\, i,j=1,2 .
\label{Eq:gij}
\end{equation}
For $i=j$, this quantity describes the autocorrelation of the fluorescence emitted by atom $j$.
Since each atom is a two-level emitter, two photons cannot be emitted simultaneously from the same atom, giving $g_{jj}^{(2)}(0)=0$.
This is the usual single-photon antibunching of an individual two-level emitter.

In contrast, the cross correlations with $i\neq j$ characterize the correlations between photons emitted from different atoms.
Specifically, $g_{12}^{(2)}(\tau)$ describes the normalized conditional probability of detecting a photon from atom~2 after detecting a photon from atom~1, while $g_{21}^{(2)}(\tau)$ describes the reverse detection order.
At zero time delay, the cross correlation becomes
\begin{equation}
  g_{12}^{(2)}(0)
  = \frac{P_{ee}}
         {\langle \sigma_1^\dagger\sigma_1\rangle \langle \sigma_2^\dagger\sigma_2\rangle},
  \label{Eq:g12_Pee}
\end{equation}
where $P_{ee}= \langle \sigma_1^\dagger\sigma_1 \sigma_2^\dagger\sigma_2 \rangle$
is the probability that both atoms are simultaneously excited.
Thus, a large fluorescence cross correlation must be interpreted together with the double-excitation probability $P_{ee}$.
This point is important because a normalized correlation function may become large in a very weak-emission regime, whereas observable generation of correlated fluorescence photon pairs requires an appreciable population of the double-excited state.

The fluorescence photon-pair emission and the cavity-field two-photon generation originate from the same two-excitation manifold, but they are associated with different dominant components.
For the cavity-field blockade, the relevant two-excitation component is mainly the two-cavity-photon state $|2,gg\rangle$.
In contrast, the fluorescence photon pairs are associated with the double-atomic-excitation component $|0,ee\rangle$, from which the two atoms can subsequently emit correlated fluorescence photons.
Therefore, the detuning selection should still be guided by the same two-excitation manifold, but the emphasis is shifted from the weight of $|2,gg\rangle$ to that of $|0,ee\rangle$.

\begin{figure}[t]
  \centering
  \includegraphics[scale=1.05]{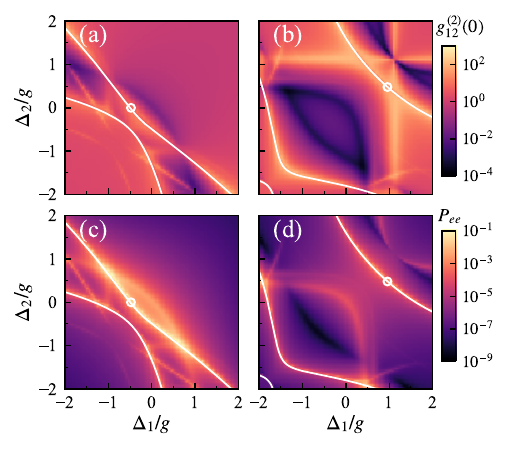}
\caption{
Detuning selection for strongly correlated fluorescence photon pairs.
Panels (a) and (b) show the equal-time fluorescence cross correlation $g_{12}^{(2)}(0)$ as a function of the atomic detunings $\Delta_1/g$ and $\Delta_2/g$, while panels (c) and (d) show the corresponding double-excitation probability $P_{ee}$.
Panels (a) and (c) are plotted for $\Delta_a=0.5g$, whereas panels (b) and (d) are plotted for $\Delta_a=1.8g$.
The white curves denote the two-excitation resonance condition $\det(H_2)=0$, and the white circles mark representative working points on the corresponding resonance branches.
Other parameters are $\kappa=\gamma_1=\gamma_2=0.01g$, $\Omega=0.04g$, $r=1$, and $\phi=\pi$.
}
\label{Fig8}
\end{figure}

\begin{figure*}[t]
\centering
\includegraphics[width=\textwidth]{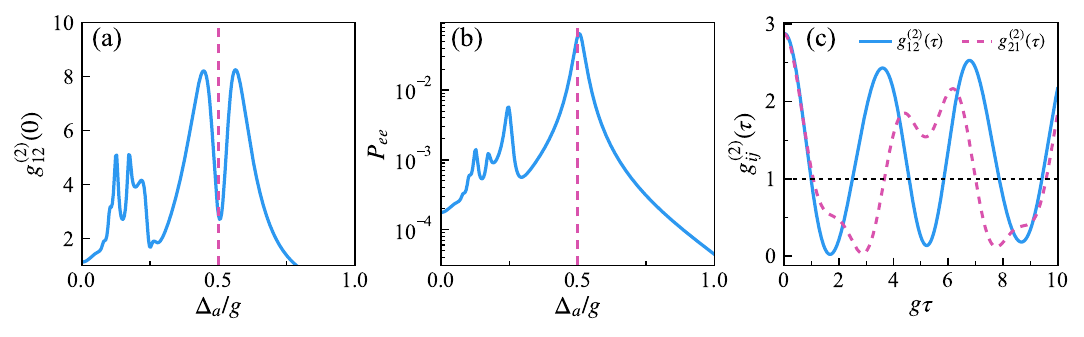}
\caption{
(a) Equal-time fluorescence cross correlation $g_{12}^{(2)}(0)$ and (b) double-excitation probability $P_{ee}$ as functions of the cavity detuning $\Delta_a/g$, with the atomic detunings fixed at $\Delta_1=-0.5g$ and $\Delta_2=0$.
The vertical dashed lines indicate the selected operating point $\Delta_a=0.5g$.
(c) Delayed fluorescence cross correlations $g_{12}^{(2)}(\tau)$ and $g_{21}^{(2)}(\tau)$ evaluated at the selected working point $\Delta_a=0.5g$, $\Delta_1=-0.5g$, and $\Delta_2=0$.
The horizontal dotted line denotes the uncorrelated value $g_{ij}^{(2)}(\tau)=1$.
Other parameters are the same as in Fig.~\ref{Fig8}.
}
\label{Fig9}
\end{figure*}

Using the two-excitation Hamiltonian $H_2$ introduced in Sec.~\ref{sec3}, we locate the resonant excitation of the two-excitation manifold from the condition $\det(H_2)=0$.
This condition does not specify a unique operating frequency, but rather defines a set of resonant parameter combinations in the three-dimensional detuning space $(\Delta_a,\Delta_1,\Delta_2)$.
Thus, for different fixed values of the cavity detuning $\Delta_a$, the two-excitation resonance can still be satisfied by properly tuning the atomic detunings $\Delta_1$ and $\Delta_2$.
However, the resonance condition alone does not determine whether a given resonant state is suitable for fluorescence photon-pair generation.
A useful operating point should exhibit both a clear fluorescence cross correlation and an appreciable population of the double-atomic-excited state.
Therefore, for each point in the $(\Delta_1,\Delta_2)$ plane, we calculate both the equal-time fluorescence cross correlation $g_{12}^{(2)}(0)$ and the double-excitation probability $P_{ee}$.

Figure~\ref{Fig8} shows two representative cavity-detuning cuts, $\Delta_a=0.5g$ and $\Delta_a=1.8g$.
For both cuts, the white curves given by $\det(H_2)=0$ agree well with the main resonance structures in the maps, confirming that the correlated fluorescence emission is closely related to the resonant excitation of the two-excitation manifold.
More importantly, the distributions of $g_{12}^{(2)}(0)$ and $P_{ee}$ reveal different physical information.
In the $\Delta_a=0.5g$ cut, the diagonal structure near the resonance curve appears as a relatively dark stripe in Fig.~\ref{Fig8}(a), while the corresponding region is bright in Fig.~\ref{Fig8}(c).
%This indicates that the double-excitation probability can be strongly enhanced even when the normalized cross correlation is not maximal.
Conversely, in the $\Delta_a=1.8g$ cut, Fig.~\ref{Fig8}(b) shows two nearly vertical bright bands with large $g_{12}^{(2)}(0)$, whereas no corresponding bright regions appear in the $P_{ee}$ map in Fig.~\ref{Fig8}(d).
%This shows that a large normalized cross correlation may arise in a weak-emission regime and does not necessarily imply an appreciable population of the double-excited state.
Therefore, the working point should not be selected solely by maximizing $g_{12}^{(2)}(0)$; it must also have an appreciable double-excitation probability.
Following this combined criterion, we choose the representative region around the white circle in the $\Delta_a=0.5g$ cut and further verify the corresponding operating detuning in Fig.~\ref{Fig9}.

To further verify the representative working point selected from Fig.~\ref{Fig8}, we fix the atomic detunings at $\Delta_1=-0.5g$ and $\Delta_2=0$ and scan the cavity detuning $\Delta_a$.
The results are shown in Fig.~\ref{Fig9}.
As shown in Fig.~\ref{Fig9}(a), the equal-time fluorescence cross correlation $g_{12}^{(2)}(0)$ remains larger than unity around $\Delta_a=0.5g$, indicating bunching between the two fluorescence photons.
However, this point corresponds to a local bunching valley rather than the maximum of the normalized cross correlation.
In contrast, Fig.~\ref{Fig9}(b) shows that the double-excitation probability $P_{ee}$ reaches a pronounced maximum at the same detuning.
This contrast provides a clear verification of the criterion discussed above: the operating point should not be chosen solely by maximizing the normalized correlation function, because the maximum of $g_{12}^{(2)}(0)$ may occur in a weaker-emission regime.
Instead, the selected point should also have an appreciable double-excitation probability.
We therefore choose $\Delta_a=0.5g$ as the operating point, where the double-atomic-excitation state is strongly populated while the fluorescence cross correlation remains above unity.

The delayed fluorescence correlations at this operating point are shown in Fig.~\ref{Fig9}(c).
Both $g_{12}^{(2)}(\tau)$ and $g_{21}^{(2)}(\tau)$ exhibit pronounced short-time values above the uncorrelated value $g_{ij}^{(2)}(\tau)=1$.
This means that the detection of a fluorescence photon from one atom enhances the probability of detecting a photon from the other atom at subsequent delay times.
The oscillatory behavior originates from the coherent atom-cavity evolution after the first photon detection.
These results demonstrate that the resonant population of the double-atomic-excitation state can be converted into strongly correlated fluorescence photon pairs.

Thus, selecting a resonant two-excitation state with an appreciable $|0, ee\rangle$ component enables strongly correlated fluorescence photon-pair emission. Together with the cavity-field source, this demonstrates two distinct two-photon outputs from different components of the same two-excitation manifold.

%In summary, the source-oriented component-selection strategy can be extended from cavity-field two-photon generation to strongly correlated fluorescence photon-pair generation within the same two-excitation manifold. In this case, the target component is the double-atomic-excitation state $|0,ee\rangle$, so that the same resonant manifold is tailored toward a different quantum-light output by selecting a different internal component. By identifying resonant operating points with appreciable $P_{ee}$ and confirming the photon-pair character through the time-delayed cross correlations, we demonstrate strongly correlated fluorescence photon-pair emission from the two atoms.

Finally, we briefly discuss the experimental feasibility of the proposed scheme.
A suitable platform is superconducting circuit QED, where two frequency-tunable artificial atoms can be coupled to a common microwave resonator and driven by phase-locked microwave fields \cite{Wallraff2004,Blais2021}.
In such systems, atom--cavity coupling strengths in the range of tens to hundreds of megahertz are experimentally accessible, while cavity and atomic decay rates in the megahertz range can be achieved with current devices.
The two driving fields can be derived from the same microwave source, so that their frequency is identical, and their amplitudes and relative phase can be calibrated independently.
Possible microwave crosstalk mainly modifies the effective values of $\Omega_1$, $\Omega_2$, and $\phi$, and can be compensated by standard microwave calibration.

To connect with experimentally realistic parameters, we take $g/2\pi=50~{\rm MHz}$. The dissipation rates used in the main calculations, $\kappa=\gamma_1=\gamma_2=0.01g$, correspond to $0.5~{\rm MHz}$. The selected detunings for the cavity-field source, $\Delta_a=0.79g$, $\Delta_1=2.09g$, and $\Delta_2=2.30g$, correspond to $39.5$, $104.5$, and $115~{\rm MHz}$, respectively. The representative driving strengths are $\Omega/2\pi\simeq6.3~{\rm MHz}$ for the cavity-field source and $2~{\rm MHz}$ for the fluorescence photon-pair source. The two atomic drives can be derived from the same microwave source, allowing their relative phase $\phi=\pi$ to be precisely controlled. Moreover, the inset of Fig.~\ref{Fig2}(a) shows that the two-photon-blockade criterion remains satisfied even at a larger dissipation rate of $0.1g$, corresponding to $5~{\rm MHz}$. Cavity-field correlations can be measured using established microwave correlation techniques \cite{Bozyigit2011,Lang2011}, while fluorescence photon-pair detection requires distinguishable atomic emission channels, for which related microwave correlation measurements have been demonstrated \cite{Gasparinetti2017}.

\section{Conclusion}
\label{sec5}
We have proposed a source-oriented component-selection strategy for two-photon quantum-light generation in a cavity-coupled two-atom system. The detunings tailor the composition of the resonant two-excitation state, while the relative driving phase controls the interference between excitation pathways. For the cavity-field source, the strategy suppresses both single- and three-photon backgrounds at matched two-photon populations, although a residual single-photon contribution remains because of the structural tradeoff between the photonic character of the target state and the spectral anharmonicity required for photon-number-selective excitation. At another working point, selecting the double-atomic-excitation component of the same two-excitation manifold generates strongly correlated fluorescence photon pairs. These results demonstrate component-selective engineering of distinct two-photon quantum-light outputs within a common light–matter platform.

\section*{Acknowledgments}
This work is supported by the National Natural Science Foundation of China (Grants Nos. 12274164, 61875067, 12375011 and 12304392) and the Fundamental Research Funds for the Central Universities (Grant No. XJ2026002801).


\begin{thebibliography}{99}

\bibitem{Kimble1977}
H. J. Kimble, M. Dagenais, and L. Mandel,
Photon antibunching in resonance fluorescence,
Phys. Rev. Lett. \textbf{39}, 691 (1977).

\bibitem{Leonski1994}
W. Leonski and R. Tanas,
Possibility of producing the one-photon state in a kicked cavity with a nonlinear Kerr medium,
Phys. Rev. A \textbf{49}, R20 (1994).

\bibitem{Carmichael1991}
H. J. Carmichael, R. J. Brecha, and P. R. Rice,
Quantum interference and collapse of the wavefunction in cavity QED,
Opt. Commun. \textbf{82}, 73 (1991).

\bibitem{Kimble2008}
H. J. Kimble,
The quantum internet,
Nature (London) \textbf{453}, 1023 (2008).

\bibitem{Gisin2007}
N. Gisin and R. Thew,
Quantum communication,
Nat. Photonics \textbf{1}, 165 (2007).

\bibitem{Ladd2010}
T. D. Ladd, F. Jelezko, R. Laflamme, Y. Nakamura, C. Monroe, and J. L. O'Brien,
Quantum computers,
Nature (London) \textbf{464}, 45 (2010).

\bibitem{Giovannetti2011}
V. Giovannetti, S. Lloyd, and L. Maccone,
Advances in quantum metrology,
Nat. Photonics \textbf{5}, 222 (2011).

\bibitem{Tian1992}
L. Tian and H. J. Carmichael,
Quantum trajectory simulations of two-state behavior in an optical cavity containing one atom,
Phys. Rev. A \textbf{46}, R6801 (1992).

\bibitem{Imamoglu1997}
A. Imamoglu, H. Schmidt, G. Woods, and M. Deutsch,
Strongly interacting photons in a nonlinear cavity,
Phys. Rev. Lett. \textbf{79}, 1467 (1997).

\bibitem{Werner1999}
M. J. Werner and A. Imamoglu,
Photon-photon interactions in cavity electromagnetically induced transparency,
Phys. Rev. A \textbf{61}, 011801(R) (1999).

\bibitem{Rebic1999}
S. Rebic, S. M. Tan, A. S. Parkins, and D. F. Walls,
Large Kerr nonlinearity with a single atom,
J. Opt. B \textbf{1}, 490 (1999).

\bibitem{Rebic2002}
S. Rebic, A. S. Parkins, and S. M. Tan,
Photon statistics of a single-atom intracavity system involving electromagnetically induced transparency,
Phys. Rev. A \textbf{65}, 063804 (2002).

\bibitem{Birnbaum2005}
K. M. Birnbaum, A. Boca, R. Miller, A. D. Boozer, T. E. Northup, and H. J. Kimble,
Photon blockade in an optical cavity with one trapped atom,
Nature (London) \textbf{436}, 87 (2005).

\bibitem{Schuster2008}
I. Schuster, A. Kubanek, A. Fuhrmanek, T. Puppe, P. W. H. Pinkse, K. Murr, and G. Rempe,
Nonlinear spectroscopy of photons bound to one atom,
Nat. Phys. \textbf{4}, 382 (2008).

\bibitem{Faraon2008}
A. Faraon, I. Fushman, D. Englund, N. Stoltz, P. Petroff, and J. Vuckovic,
Coherent generation of nonclassical light on a chip via photon-induced tunneling and blockade,
Nat. Phys. \textbf{4}, 859 (2008).

\bibitem{Reinhard2012}
A. Reinhard, T. Volz, M. Winger, A. Badolato, K. J. Hennessy, E. L. Hu, and A. Imamoglu,
Strongly correlated photons on a chip,
Nat. Photonics \textbf{6}, 93 (2012).

\bibitem{Kyriienko2020}
O. Kyriienko, D. N. Krizhanovskii, and I. A. Shelykh,
Nonlinear quantum optics with trion polaritons in 2D monolayers: Conventional and unconventional photon blockade,
Phys. Rev. Lett. \textbf{125}, 197402 (2020).

\bibitem{Lang2011}
C. Lang, D. Bozyigit, C. Eichler, L. Steffen, J. M. Fink, A. A. Abdumalikov Jr., M. Baur, S. Filipp, M. P. da Silva, A. Blais, and A. Wallraff,
Observation of resonant photon blockade at microwave frequencies using correlation function measurements,
Phys. Rev. Lett. \textbf{106}, 243601 (2011).

\bibitem{Hoffman2011}
A. J. Hoffman, S. J. Srinivasan, S. Schmidt, L. Spietz, J. Aumentado, H. E. Tureci, and A. A. Houck,
Dispersive photon blockade in a superconducting circuit,
Phys. Rev. Lett. \textbf{107}, 053602 (2011).

\bibitem{Bozyigit2011}
D. Bozyigit, C. Lang, L. Steffen, J. M. Fink, C. Eichler, M. Baur, R. Bianchetti, P. J. Leek, S. Filipp, M. P. da Silva, A. Blais, and A. Wallraff,
Antibunching of microwave-frequency photons observed in correlation measurements using linear detectors,
Nat. Phys. \textbf{7}, 154 (2011).

\bibitem{Chakram2022}
S. Chakram, K. He, A. V. Dixit, A. E. Oriani, R. K. Naik, N. Leung, H. Kwon, W.-L. Ma, L. Jiang, and D. I. Schuster,
Multimode photon blockade,
Nat. Phys. \textbf{18}, 879 (2022).

\bibitem{Trivedi2019}
R. Trivedi, M. Radulaski, K. A. Fischer, S. Fan, and J. Vuckovic,
Photon blockade in weakly driven cavity quantum electrodynamics systems with many emitters,
Phys. Rev. Lett. \textbf{122}, 243602 (2019).

\bibitem{Chen2022}
M. Chen, J. Tang, L. Tang, H. Wu, and K. Xia,
Photon blockade and single-photon generation with multiple quantum emitters,
Phys. Rev. Res. \textbf{4}, 033083 (2022).

\bibitem{Liew2010}
T. C. H. Liew and V. Savona,
Single photons from coupled quantum modes,
Phys. Rev. Lett. \textbf{104}, 183601 (2010).

\bibitem{Bamba2011}
M. Bamba, A. Imamoglu, I. Carusotto, and C. Ciuti,
Origin of strong photon antibunching in weakly nonlinear photonic molecules,
Phys. Rev. A \textbf{83}, 021802(R) (2011).

\bibitem{Flayac2017}
H. Flayac and V. Savona,
Unconventional photon blockade,
Phys. Rev. A \textbf{96}, 053810 (2017).

\bibitem{Snijders2018}
H. J. Snijders, J. A. Frey, J. Norman, H. Flayac, V. Savona, A. C. Gossard, J. E. Bowers, M. P. van Exter, D. Bouwmeester, and W. Loffler,
Observation of the unconventional photon blockade,
Phys. Rev. Lett. \textbf{121}, 043601 (2018).

\bibitem{Vaneph2018}
C. Vaneph, A. Morvan, G. Aiello, M. Fechant, M. Aprili, J. Gabelli, and J. Esteve,
Observation of the unconventional photon blockade in the microwave domain,
Phys. Rev. Lett. \textbf{121}, 043602 (2018).

\bibitem{Zubizarreta2020}
E. Zubizarreta Casalengua, J. C. Lopez Carreno, F. P. Laussy, and E. del Valle,
Conventional and unconventional photon statistics,
Laser Photonics Rev. \textbf{14}, 1900279 (2020).

\bibitem{Hou2019}
K. Hou, C. J. Zhu, Y. P. Yang, and G. S. Agarwal,
Interfering pathways for photon blockade in cavity QED with one and two qubits,
Phys. Rev. A \textbf{100}, 063817 (2019).

\bibitem{Miranowicz2013}
A. Miranowicz, M. Paprzycka, Y.-X. Liu, J. Bajer, and F. Nori,
Two-photon and three-photon blockades in driven nonlinear systems,
Phys. Rev. A \textbf{87}, 023809 (2013).

\bibitem{Hovsepyan2014}
G. H. Hovsepyan, A. R. Shahinyan, and G. Y. Kryuchkyan,
Multiphoton blockades in pulsed regimes beyond stationary limits,
Phys. Rev. A \textbf{90}, 013839 (2014).

\bibitem{Deng2015}
W.-W. Deng, G.-X. Li, and H. Qin,
Enhancement of the two-photon blockade in a strong-coupling qubit-cavity system,
Phys. Rev. A \textbf{91}, 043831 (2015).

\bibitem{Hamsen2017}
C. Hamsen, K. N. Tolazzi, T. Wilk, and G. Rempe,
Two-photon blockade in an atom-driven cavity QED system,
Phys. Rev. Lett. \textbf{118}, 133604 (2017).

\bibitem{Felicetti2018}
S. Felicetti, D. Z. Rossatto, E. Rico, E. Solano, and P. Forn-Diaz,
Two-photon quantum Rabi model with superconducting circuits,
Phys. Rev. A \textbf{97}, 013851 (2018).

\bibitem{VillasBoas2019}
C. J. Villas-Boas and D. Z. Rossatto,
Multiphoton Jaynes-Cummings model: Arbitrary rotations in Fock space and quantum filters,
Phys. Rev. Lett. \textbf{122}, 123604 (2019).

\bibitem{Zou2020}
F. Zou, X.-Y. Zhang, X.-W. Xu, J.-F. Huang, and J.-Q. Liao,
Multiphoton blockade in the two-photon Jaynes-Cummings model,
Phys. Rev. A \textbf{102}, 053710 (2020).


\bibitem{Li2024TJC}
H.-J. Li, L.-B. Fan, S. Ma, J.-Q. Liao, and C.-C. Shu,
Exploring photon blockade in a two-photon Jaynes-Cummings model with atom and cavity drivings,
Phys. Rev. A \textbf{110}, 043707 (2024).


\bibitem{Zhu2017}
C. J. Zhu, Y. P. Yang, and G. S. Agarwal,
Collective multiphoton blockade in cavity quantum electrodynamics,
Phys. Rev. A \textbf{95}, 063842 (2017).

\bibitem{Bin2018}
Q. Bin, X.-Y. Lu, S.-W. Bin, and Y. Wu,
Two-photon blockade in a cascaded cavity-quantum-electrodynamics system,
Phys. Rev. A \textbf{98}, 043858 (2018).

\bibitem{Lin2019}
J. Z. Lin, K. Hou, C. J. Zhu, and Y. P. Yang,
Manipulation and improvement of multiphoton blockade in a cavity-QED system with two cascade three-level atoms,
Phys. Rev. A \textbf{99}, 053850 (2019).

\bibitem{Tang2024}
J. Tang and Y. Deng,
Tunable multiphoton bundles emission in a Kerr-type two-photon Jaynes-Cummings model,
Phys. Rev. Res. \textbf{6}, 033247 (2024).

\bibitem{Zhang2025}
G.-Y. Zhang, Z.-H. Liu, J.-Q. Liao, and X.-W. Xu,
Multiphoton blockade by frequency-matched multitone drive,
Phys. Rev. A \textbf{112}, 053703 (2025).


\bibitem{Qiao2024}
X. Qiao, Z. Yao, and H. Yang,
Strongly enhanced photon-pair blockade with three-wave mixing by quantum interference,
Phys. Rev. A \textbf{110}, 053702 (2024).

\bibitem{Kowalewska2019}
A. Kowalewska-Kudlaszyk, S. I. Abo, G. Chimczak, J. Perina Jr., F. Nori, and A. Miranowicz,
Two-photon blockade and photon-induced tunneling generated by squeezing,
Phys. Rev. A \textbf{100}, 053857 (2019).

\bibitem{Feng2021}
L.-J. Feng and S.-Q. Gong,
Two-photon blockade generated and enhanced by mechanical squeezing,
Phys. Rev. A \textbf{103}, 043509 (2021).

\bibitem{Li2024FW}
Y. Li, Z. Yao, and H. Yang,
One-photon and two-photon blockades in a four-wave-mixing system embedded with an atom,
Phys. Rev. A \textbf{109}, 043702 (2024).

\bibitem{Lin2024}
H. Lin, X. Luo, X. Wang, F. Gao, Y. Zhou, and Z. Yao,
Inducing tunable conventional photon blockade and two-photon blockade in a second-order nonlinear system with two-level atoms,
Opt. Express \textbf{32}, 23056 (2024).

\bibitem{Rabl2011}
P. Rabl,
Photon blockade effect in optomechanical systems,
Phys. Rev. Lett. \textbf{107}, 063601 (2011).

\bibitem{Nunnenkamp2011}
A. Nunnenkamp, K. Borkje, and S. M. Girvin,
Single-photon optomechanics,
Phys. Rev. Lett. \textbf{107}, 063602 (2011).

\bibitem{Xu2013}
X.-W. Xu, Y.-J. Li, and Y.-X. Liu,
Photon-induced tunneling in optomechanical systems,
Phys. Rev. A \textbf{87}, 025803 (2013).

\bibitem{Liao2013}
J.-Q. Liao and F. Nori,
Photon blockade in quadratically coupled optomechanical systems,
Phys. Rev. A \textbf{88}, 023853 (2013).

\bibitem{Komar2013}
P. Komar, S. D. Bennett, K. Stannigel, S. J. M. Habraken, P. Rabl, P. Zoller, and M. D. Lukin,
Single-photon nonlinearities in two-mode optomechanics,
Phys. Rev. A \textbf{87}, 013839 (2013).

\bibitem{Xie2016}
H. Xie, G. W. Lin, X. Chen, Z. H. Chen, and X. M. Lin,
Single-photon nonlinearities in a strongly driven optomechanical system with quadratic coupling,
Phys. Rev. A \textbf{93}, 063860 (2016).

\bibitem{Solki2023}
H. Solki, A. Motazedifard, and M. H. Naderi,
Improving photon blockade, entanglement, and mechanical-cat-state generation in a generalized cross-Kerr optomechanical circuit,
Phys. Rev. A \textbf{108}, 063505 (2023).

\bibitem{Tian2025Disorder}
G. Tian, L.-L. Zheng, Z.-M. Zhan, F. Nori, and X.-Y. L\"u,
Disorder-induced strongly correlated photons in waveguide QED,
Phys. Rev. Lett. \textbf{135}, 153604 (2025).


\bibitem{Huang2018}
R. Huang, A. Miranowicz, J.-Q. Liao, F. Nori, and H. Jing,
Nonreciprocal photon blockade,
Phys. Rev. Lett. \textbf{121}, 153601 (2018).

\bibitem{Li2019}
B. Li, R. Huang, X. Xu, A. Miranowicz, and H. Jing,
Nonreciprocal unconventional photon blockade in a spinning optomechanical system,
Photonics Res. \textbf{7}, 630 (2019).

\bibitem{Gou2023Simultaneous}
C. Gou and X. Hu,
Simultaneous nonreciprocal photon blockade in two coupled spinning resonators via Sagnac-Fizeau shift and parametric amplification,
Phys. Rev. A \textbf{108}, 043723 (2023).

\bibitem{Lu2024Chiral}
Z.-G. Lu, Y. Wu, and X.-Y. L\"u,
Chiral interaction induced near-perfect photon blockade,
Phys. Rev. Lett. \textbf{134}, 013602 (2025).

\bibitem{Xu2018NRQuadratic}
X.-W. Xu, Y.-J. Zhao, H. Wang, H. Jing, and A.-X. Chen,
Nonreciprocal photon blockade via quadratic optomechanical coupling,
Photon. Res. \textbf{8}, 143 (2020).

\bibitem{Xu2019Synthetic}
X.-W. Xu, Y. Li, B. Li, H. Jing, and A.-X. Chen,
Nonreciprocity via nonlinearity and synthetic magnetism,
Phys. Rev. Applied \textbf{13}, 044070 (2020).

\bibitem{Li2024Loss}
B. Li, Y. Zuo, L.-M. Kuang, H. Jing, and C. Lee,
Loss-induced quantum nonreciprocity,
npj Quantum Inf. \textbf{10}, 75 (2024).

\bibitem{Sun2023}
J. Y. Sun and H. Z. Shen,
Photon blockade in non-Hermitian optomechanical systems with nonreciprocal couplings,
Phys. Rev. A \textbf{107}, 043715 (2023).

\bibitem{Huang2022EP}
R. Huang, S. K. \"Ozdemir, J.-Q. Liao, F. Minganti, L.-M. Kuang, F. Nori, and H. Jing,
Exceptional photon blockade: Engineering photon blockade with chiral exceptional points,
Laser Photonics Rev. \textbf{16}, 2100430 (2022).

\bibitem{Geng2024}
Z. Geng, Y. Chen, Y. Jiang, Y. Xia, and J. Song,
Engineering dynamical photon blockade with Liouville exceptional points,
Opt. Lett. \textbf{49}, 3026 (2024).

\bibitem{Ghosh2019}
S. Ghosh and T. C. H. Liew,
Dynamical blockade in a single-mode bosonic system,
Phys. Rev. Lett. \textbf{123}, 013602 (2019).

\bibitem{Li2022}
M. Li, Y.-L. Zhang, S.-H. Wu, C.-H. Dong, X.-B. Zou, G.-C. Guo, and C.-L. Zou,
Single-mode photon blockade enhanced by bi-tone drive,
Phys. Rev. Lett. \textbf{129}, 043601 (2022).

\bibitem{Zhang2024}
G.-Y. Zhang, Z.-H. Liu, and X.-W. Xu,
Optimizing dynamical blockade via a particle-swarm-optimization algorithm,
Phys. Rev. A \textbf{110}, 023718 (2024).


\bibitem{Zhu1987}
S.-Y. Zhu and X.-S. Li, Quantum theory of a two-photon laser, Phys. Rev. A \textbf{36}, 3889 (1987).

\bibitem{Lewenstein1990}
M. Lewenstein, Y. Zhu, and T. W. Mossberg, Two-photon gain and lasing in strongly driven two-level atoms, Phys. Rev. Lett. \textbf{64}, 3131 (1990).

\bibitem{Kubanek2008}
A. Kubanek, A. Ourjoumtsev, I. Schuster, M. Koch, P. W. H. Pinkse, K. Murr, and G. Rempe, Two-photon gateway in one-atom cavity quantum electrodynamics, Phys. Rev. Lett. \textbf{101}, 203602 (2008).


\bibitem{SanchezMunoz2014}
C. S\'anchez Mu\~noz, E. del Valle, A. Gonz\'alez-Tudela, K. M\"uller,
S. Lichtmannecker, M. Kaniber, C. Tejedor, J. J. Finley, and F. P. Laussy,
Emitters of $N$-photon bundles,
Nat. Photonics \textbf{8}, 550 (2014).


%===================================================================================================

\bibitem{Wallraff2004}
A. Wallraff, D. I. Schuster, A. Blais, L. Frunzio, R. -S. Huang, J. Majer, S. Kumar, S. M. Girvin, and R. J. Schoelkopf,
Strong coupling of a single photon to a superconducting qubit using circuit quantum electrodynamics,
Nature \textbf{431}, 162 (2004).

\bibitem{Blais2021}
A. Blais, A. L. Grimsmo, S. M. Girvin, and A. Wallraff,
Circuit quantum electrodynamics,
Rev. Mod. Phys. \textbf{93}, 025005 (2021).

\bibitem{Gasparinetti2017}
S. Gasparinetti, M. Pechal, J.-C. Besse, M. Mondal, C. Eichler, and A. Wallraff,
Correlations and entanglement of microwave photons emitted in a cascade decay,
Phys. Rev. Lett. \textbf{119}, 140504 (2017).
\end{thebibliography}
\end{document}